\documentclass[british]{heart2013}
\usepackage[T1]{fontenc}
\usepackage[latin9]{inputenc}
\usepackage{color}
\usepackage{float}
\usepackage{graphicx}

\makeatletter

\floatstyle{ruled}
\newfloat{algorithm}{tbp}{loa}
\providecommand{\algorithmname}{Algorithm}
\floatname{algorithm}{\protect\algorithmname}

\newenvironment{lyxcode}
{\par\begin{list}{}{
\setlength{\rightmargin}{\leftmargin}
\setlength{\listparindent}{0pt}
\raggedright
\setlength{\itemsep}{0pt}
\setlength{\parsep}{0pt}
\normalfont\ttfamily}%
 \item[]}
{\end{list}}

\makeatother

\begin{document}
\title{Cache-aware Parallel Programming for Manycore Processors}

\author{Ashkan Tousimojarad\\
	\normalsize School of Computing Science\\
	\normalsize University of Glasgow, Glasgow, UK\\
	\large a.tousimojarad.1@research.gla.ac.uk 
\and 
	Wim Vanderbauwhede \\
	\normalsize School of Computing Science\\
	\normalsize University of Glasgow, Glasgow, UK\\
	\large wim@dcs.gla.ac.uk}

\maketitle
\begin{abstract}
With rapidly evolving technology, multicore and manycore processors
have emerged as promising architectures to benefit from increasing
transistor numbers. The transition towards these parallel architectures
makes today an exciting time to investigate challenges in parallel
computing. The TILEPro64 is a manycore accelerator, composed of 64
tiles interconnected via multiple 8$\times$8 mesh networks. It contains
per-tile caches and supports cache-coherent shared memory by default.
In this paper we present a programming technique to take advantages
of distributed caching facilities in manycore processors. However,
unlike other work in this area, our approach does not use architecture-specific
libraries. Instead, we provide the programmer with a novel technique
on how to program future \textcolor{black}{Non-Uniform Cache Architecture}
(NUCA) manycore systems, bearing in mind their caching organisation.
We show that our \emph{localised} programming approach can result
in a significant improvement of the parallelisation efficiency (speed-up).
\end{abstract}

\section{Introduction}

Modern processors provide hierarchical caching. To achieve optimal
performance, manycore programmers need to know about the target platform
caching organisation as much as they need to know about how to parallelise
their applications. Designs with larger number of cores tend towards
scalable tiled architectures with physically distributed shared caches.
The asymmetry in the physical distances between cores results in a
design called \textcolor{black}{Non-Uniform Cache Architecture (NUCA),
in which a home core (tile) is associated with each memory address,
and the access latency to the home core depends on the physical on-die
location of the requesting core} \cite{kim2002adaptive}\cite{shim2012judicious}.

In this paper, we present a programming technique to utilise manycores
with per-core caches. We validate our approach with a parallel sorting
algorithm. 

Sorting algorithms have attracted a great attention, due to their
simple concept but complex optimisation. They are basic blocks of
majority of applications, such as databases, data mining applications,
and computer graphics \cite{morari2012efficient}. Furthermore, they
are memory-bound, which makes them suitable candidates for investigating
the effect of memory organisation on performance. Although based on
their time complexity, most of them do not scale linearly with the
number of threads, they can still be parallelised easily and effectively. 

Merge sort is an efficient divide-and-conquer algorithm and its parallel
version is a great example of parallel reduction. The average complexity
of its serial version is \textit{O(n log n)}. We use parallel merge
sort as our test case to demonstrate a programming approach that can
be applied to many other parallel algorithms. We also discuss that
the performance of parallel applications on manycores cannot be estimated
only based on their time complexity. The memory architecture plays
an important role. Moreover, as the number of cores grows, memory
contention becomes increasingly significant.

For the purposes of this research, we used the TILEPro64 Tile Processor
from Tilera \cite{bell2008tile64}. This manycore defines a globally
shared, flat 36-bit physical address space and a 32-bit virtual address
space. The global address space allows for instructions and data sharing
between processes and threads. There is a large difference between
the DRAM access time and the speed of the cores, which makes the cache
organisation a crucial part of this architecture. Tilera's cache organisation
-- called Dynamic Distributed Cache (DDC) -- is flexible and software
configurable. Its aim is to provide a hardware-managed, cache-coherent
approach to shared memory. DDC allows a page of the shared memory
to be homed on a single tile or hashed across a set of tile. Other
tiles can cache this page remotely.

\section{The TILEPro64 }

The Tilera TILEPro64 Tile Processor is a manycore accelerator composed
of 64 tiles interconnected via multiple 8$\times$8 mesh networks.
It provides distributed cache-coherent shared memory by default. It
has 16GB of DDR memory, but in order to use the global address space
shared among all tiles, addressing is limited to 32-bit, i.e. 4GB. 

Each physical memory address in the TILEPro64 is associated with a
\textit{home tile}. Cache coherent view of the memory which is a key
requirement in shared memory programming models is served through
the \textit{home tile}. A hardware coherency mechanism is also used
to guarantee cache coherence among the tiles. Therefore, it is possible
to cache read-write regions of memory in the cache of the tile running
the code (local cache). The copy of each cache line can be requested
from its \textit{home tile}. If another tile writes new data to the
cache line, the \textit{home tile} is responsible to invalidate all
copies, and other tiles have to refetch the newer version. This behaviour
makes DDC a dynamic cache organisation. Caching the data by the \textit{home
tile} itself is called L3 cache, because the home tile can be thought
of as a higher level beyond the L2 cache. In other words, this concept
can be thought of as having a virtual L3 cache on top of the actual
local L2 caches. If an L2 miss occurs, the request will be first sent
to\textit{ home tile} rather than directly to the DDR memory. Thus,
the distributed L3 cache comprises the union of all L2 caches. 

Another feature of DDC, called \textit{hash for home}, is the capability
of distributing the \textit{home cache} of memory regions between
different tiles at a cache-line granularity. As a result, the potential
for hot spots is reduced, and the request traffic will be distributed
across the whole chip.

The homing mechanism is basically intended to provide cache coherence,
though it can also improve the performance by reducing the read instruction
latencies. There are three different classes of homing in the Tile
Processor system: I) Local homing, II) Remote homing, III) Hash for
home. The local homing strategy homes the entire memory page on the
same tile that is accessing the memory. Therefore, on an L2 miss,
a request is sent directly to DDR memory. With the remote homing,
a different tile than the one accessing the memory is used to home
the entire memory page. Therefore, on an L2 miss, a request is first
sent to the remote home tile's cache (which can be called the L3 cache).
The \textit{hash for home} strategy as described above is a new feature
of DDC, which is very similar to remote homing. The only difference
is that instead of mapping an entire memory page to a single home
tile, it is hashed across different tiles at a cache-line granularity.

The Tilera's version of Symmetric Multiprocessing (SMP) Linux, called
Tile Linux, which is based on the standard open-source Linux version
2.6.26, by default sets the home cache for a given page to be \textit{hash
for home}. This can be changed by the \textit{ucache\_hash} boot option.
In this work, we demonstrate how our programming approach allows to
leverage possible options provided by the hypervisor. The main reason
behind our study is that we believe the \textit{hash for home} policy
at the cache line granularity is too fine-grained. Applying our technique,
the programs not only maintain their performance under the \textit{hash
for home} policy, but also show better performance under the \textit{local
homing} policy. In our approach, the data chunks, distributed across
the chip. have the size of\textit{ input\_size/num\_threads}, and
each chunk is homed on the tile on which its task is running.

\section{Related Work}

Data locality in NUCA designs is discussed in \cite{shim2012judicious}.
The authors propose an on-line prediction strategy which decides whether
to perform a remote access (as in traditional NUCA designs) or to
migrate a thread at the instruction level. In \cite{muddukrishna2013task},
the NUCA characterisation of the TILEPro64 is explored. Based on this
characterisation, a home cache aware task scheduling is developed
to distribute task data on home caches. Although one of the aims of
this work is to provide simple interfaces, still its memory allocation
policy is a wrapper around the architecture-specific API, which makes
the code dependent to the platform. A similar work to ours on sorting
but with different methods and purposes is performed on the TILEPro64
\cite{morari2012efficient}. They have targeted throughput and power
efficiency of the radix sort algorithm employing fine-grained control
and various optimisation techniques offered by the Tilera Multicore
Components (TMC) API. The idea of the conventional recursive parallel
merge sort in OpenMP is borrowed from \cite{radenskishared}. In our
previous work \cite{tousimojarad2013parallel}, we have shown how
this algorithm scales on the TILEPro64 compared to the theoretical
model.

\section{Methodology}

We introduce our \textit{localisation} approach for memory-bound array
computations which is based upon three building blocks: I) Local homing
caching strategy, II) Static thread mapping, III) Home cache localisation.
By disabling the \textit{hash for home} strategy, we guarantee that
all types of user memory are locally homed on the task's current tile.
The second step is to statically map each thread to a processing core
(tile). We will show that leaving the decision of the thread mapping
to the Tile Linux can be costly. The Tile Linux tries to migrate the
threads during the execution time, and those migrations are costly
not only in terms of cache misses but also because of the resulting
delay. The final step is the key in our approach and its job is to
localise the home caches. With the local homing strategy, the whole
array is homed on the tile on which it has been created. Instead,
if each working thread copies its corresponding part of the array
into a new dynamic array, then every newly created array will be homed
on the same tile as the one its thread is mapped to.

As mentioned above, it is possible to choose a single home cache for
a given memory page rather than hash it across a set of tiles (all
of the tiles, by default). Although the idea behind the \textit{hash
for home} policy is to decentralise the home cache and to reduce the
hot spots, sequential accesses cannot benefit from this fine-grained
policy. The reason is that with this policy, the sequential parts
of an array are homed on different tiles, and have different access
latencies. A comprehensive discussion is provided in \cite{morari2012efficient}.
On the other hand, with the \textit{local homing}, the array is homed
on the core on which its task in running, and obviously the whole
traffic on a single L2 cache makes it inefficient. Based on this background,
we present our method as a set of simple steps to exploit data locality
on each tile:

\begin{algorithm}[H]
\textsf{1- Divide the input array of size }\textsf{\textit{n}}\textsf{
to }\textsf{\textit{m}}\textsf{ parts, where }\textsf{\textit{m}}\textsf{
is the number of threads.}

\textsf{2- Assign each thread a part of the whole array, by passing
pointers.}

\textsf{3- Map each thread to a core.}

\textit{{*}Localisation of the home caches occurs in the next step}{*}

\textsf{4- Copy each part to a new array of size }\textsf{\textit{n/m}}\textsf{.}

\textsf{5- Free the dynamically allocated memory as soon as each thread
finishes its job.}

\caption{\label{alg:Localisation}\textit{Localisation} at a coarse granularity
for parallel array computations}
\end{algorithm}

Therefore, with this method, each dynamically created smaller array
will be homed on the tile on which its corresponding thread is running. 

For the rest of the paper, we call our new technique the \textit{localised}
approach, and call the conventional way of programming the \textit{non-localised}
approach. Although the first 2 steps of Algorithm \ref{alg:Localisation}
are common between both approaches, we have put them altogether to
be employed as a programming style in any memory-bound parallel array
computation. Two of the three building blocks of the \textit{localisation}
approach are listed in the Algorithm \ref{alg:Localisation} (steps
3 and 4). The other one can be set by the system hypervisor, such
that the single-tile homing becomes enabled. The 5th step is used
for efficient memory management. 

In order to show how these 5 steps are applied to an array computation
program, we have designed a micro-benchmark that creates two arrays:
one input and one output. It initialises the input array, and then
inside each thread a part of the input array is copied to the corresponding
part of the output array. The size of the arrays is 1 million integers.
Each thread has a loop that repeats the copying operations, and the
number of these repetitions is used for our evaluation. The array
is distributed among 63 threads (the maximum numbers of cores available
in the TILEPro64). If the number of repetitions is small, the execution
time will become so short that the difference cannot be observed.

The first 3 steps of Algorithm \ref{alg:Localisation} will be shown
in the next section inside a parallel merge sort program. For the
micro-benchmark, we only show how to localise the home caches within
each working thread. We have measured the execution time for the conventional
approach (\textit{non-localised}) under the Tile Linux default mapping
strategy, along with the \textit{hash for home} policy. For the \textit{localisation}
approach, we have disabled the \textit{hash for home}, mapped each
thread to one core, and made the changes highlighted in Algorithm
\ref{alg:micro-benchmark}.

\begin{algorithm}
\begin{lyxcode}
{\small ...}{\small \par}

//~Each~thread~gets~a~part~of~the~input~array

//~\textit{Non-localised}~Approach:

{\small repetitive\_copy(input1,output,size);}{\small \par}

//~{*}{*}{*}{*}{*}~OR~{*}{*}{*}{*}{*}

//~\textit{Localised}~Approach:

//~Create~a~local~copy~of~each~part

\textcolor{blue}{\small int{*}~input\_cpy1~=~new~int{[}size{]};~~~~~	}{\small \par}

\textcolor{blue}{\small memcpy(input\_cpy1,input1,size{*}sizeof(int));~~~~~	}{\small \par}

\textcolor{blue}{\small repetitive\_copy(input\_cpy1,~output,~size);~~~~~	}{\small \par}

\textcolor{blue}{\small free(input\_cpy1);~}{\small \par}

{\small ...	}{\small \par}
\end{lyxcode}
\caption{\label{alg:micro-benchmark}Localisation of the home caches in the
micro-benchmark }
\end{algorithm}

\begin{figure}[t]
\begin{centering}
\includegraphics[width=0.9\columnwidth]{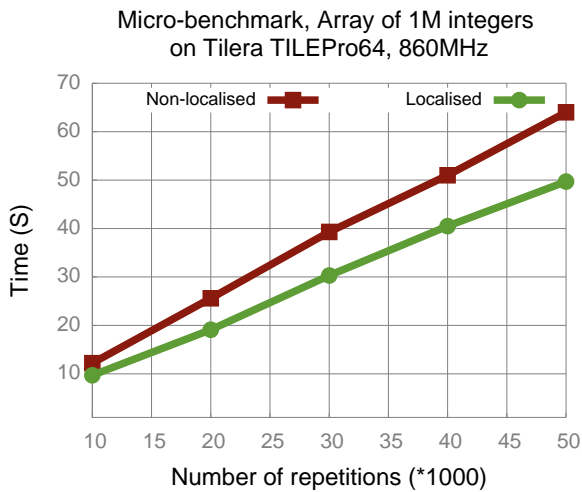}
\par\end{centering}

\caption{\label{fig:Execution_Micro}Execution time of the micro-benchmark.
Localised vs. Non-localised}
\end{figure}

The result of applying our technique to the micro-benchmark is promising
(Figure \ref{fig:Execution_Micro}). It shows more performance gain
as the number of accesses becomes larger.

For the rest of this paper, we use an OpenMP version of a recursive
parallel merge sort algorithm as a short program which can apply our
method effectively. We have used it as a memory-bound parallel algorithm
to investigate the effect of our programming approach on the execution
time in a real world example. The corresponding C++ code for the merge
sort example is shown in Algorithm \ref{alg:C++-code-example}. Although
it is possible to parallelise this algorithm further by parallelising
the merge operation, our purpose is not to optimise this particular
algorithm.

The method is straightforward and can be generally applied to any
parallelisable array computation, where each part of the array is
accessed multiple times. An example of the worst situation is the
merge sort algorithm itself, where the algorithm is not in-place,
and in each level of the reduction tree, a double sized auxiliary
array called \textit{scratch} is required for the merging operation.
Since we want to examine the effect of one parameter at a time, we
call this issue an \textit{intermediate step} and explore the effect
of it separately in Section \ref{Effect-of-Data}.

\begin{algorithm}
\begin{lyxcode}
{\small }{\small \par}

{\small int~counter~=~-1;}{\small \par}

{\small int~NUM\_CORES~=~omp\_get\_num\_procs();}{\small \par}

{\small int{*}~mergesort\_serial(int{*}~input,~}{\small \par}

{\small{}~~int{*}~scratch,~int~size)~\{}{\small \par}

{\small{}~~~~~...}{\small \par}

//~Recursive~serial~merge~sort

{\small{}~~~~~...}{\small \par}

{\small{}~~~~~return~input;}{\small \par}

{\small \}}{\small \par}

{\small int{*}~merge(int{*}~input1,~int~size1,~}{\small \par}

{\small{}~~int{*}~input2,~int~size2,~int{*}~scratch)~\{}{\small \par}

{\small{}~~~~~...}{\small \par}

{\small{}~~~~~memcpy~(input1,}{\small \par}

{\small{}~~~~~~~scratch,~(size1+size2){*}sizeof(int));}{\small \par}

{\small{}~~~~~return~input1;}{\small \par}

{\small \}}{\small \par}

{\small int{*}~mergesort\_parallel\_omp(int{*}~input,~}{\small \par}

{\small{}~~int{*}~scratch,~int~size,~int~threads)~\{}{\small \par}

{\small{}~~~~~if~(threads~==~1)~\{}{\small \par}

{\small \#pragma~omp~critical}{\small \par}

{\small{}~~~~~~~~\{}{\small \par}

{\small{}~~~~~~~~~counter++;}{\small \par}

//~Static~mapping~in~the~ordered~way

{\small \#ifdef~STATIC\_MAPPING}{\small \par}

{\small{}~~~~~~~~~cpu\_set\_t~mask;}{\small \par}

{\small{}~~~~~~~~~CPU\_ZERO(\&mask);}{\small \par}

{\small{}~~~~~~~~~int~target=counter\%NUM\_CORES;}{\small \par}

{\small{}~~~~~~~~~CPU\_SET(target,~\&mask);}{\small \par}

{\small{}~~~~~~~~~if~(sched\_setaffinity(0,~}{\small \par}

{\small{}~~~~~~~~~~~~~sizeof(mask),~\&mask)~!=~0)}{\small \par}

{\small{}~~~~~~~~~~~perror(``sched\_setaffinity'');}{\small \par}

{\small \#endif}{\small \par}

{\small{}~~~~~~~~~\}}{\small \par}

{\small{}~~~~~~~~~int{*}~r~=~mergesort\_serial(input,~}{\small \par}

{\small{}~~~~~~~~~~~~~~~~~~~~scratch,~size);}{\small \par}

{\small{}~~~~~~~~~return~r;}{\small \par}

{\small{}~~~~~\}}{\small \par}

{\small{}~~~~~else~if~(threads~>~1)~\{}{\small \par}

{\small \#pragma~omp~parallel~sections}{\small \par}

{\small{}~~~~~\{}{\small \par}

{\small \#pragma~omp~section}{\small \par}

{\small{}~~~~~~\{}{\small \par}

{\small{}~~~~~~~mergesort\_parallel\_omp(input,~}{\small \par}

{\small{}~~~~~~~~~scratch,~size/2,~threads/2);}{\small \par}

{\small{}~~~~~~\}}{\small \par}

{\small \#pragma~omp~section}{\small \par}

{\small{}~~~~~~\{}{\small \par}

{\small{}~~~~~~~mergesort\_parallel\_omp(input+size/2,~}{\small \par}

{\small{}~~~~~~~~~scratch+size/2,~size-size/2,~}{\small \par}

{\small{}~~~~~~~~~threads-threads/2);}{\small \par}

{\small{}~~~~~~\}}{\small \par}

{\small{}~~~~~\}}{\small \par}

{\small{}~~~~~return~merge(input,~size/2,~input+size/2,~}{\small \par}

{\small{}~~~~~~~~~~~~~~size-size/2,~scratch);}{\small \par}

{\small{}~~~~\}}{\small \par}

{\small{}~~return~(int{*})0;~}{\small \par}

{\small \}}{\small \par}

{\small int~main()~\{}{\small \par}

{\small{}~~~~omp\_set\_nested(1);}{\small \par}

{\small{}~~~~omp\_set\_num\_threads(2);}{\small \par}

{\small{}~~~~int{*}~array0~=~new~int{[}ARRAY\_SZ{]};}{\small \par}

{\small{}~~~~int{*}~scratch0~=~new~int{[}ARRAY\_SZ{]};}{\small \par}

{\small{}~~~~...}{\small \par}

{\small{}~~~~mergesort\_parallel\_omp(array0,~}{\small \par}

{\small{}~~~~~~scratch0,~ARRAY\_SZ,~NUM\_THREADS);}{\small \par}

{\small{}~~~~free(scratch0);}{\small \par}

{\small{}~~~~...}{\small \par}

{\small{}~~~~return~0;}{\small \par}

{\small \}}{\small \par}
\end{lyxcode}
\caption{\label{alg:C++-code-example}C++ merge sort code for the \textit{non-localised}
approach ({[}5{]})}
\end{algorithm}

As shown in Algorithm \ref{alg:Changes-to-code}, only some minor
changes are required to apply the \textit{localisation} approach to
the conventional code in order to benefit from the \textit{local homing}
policy.
\begin{algorithm}
\begin{lyxcode}
{\small ...}{\small \par}

{\small int{*}~mergesort\_serial(int{*}~input,~}{\small \par}

{\small{}~~int{*}~scratch,~int~size)~\{}{\small \par}

//~Create~local~copy~of~the~input~array

{\small{}~~~~~}\textcolor{blue}{\small int{*}~input\_cpy~=~new~int{[}size{]};}{\small \par}

\textcolor{blue}{\small{}~~~~~memcpy(input\_cpy,~input,~size{*}sizeof(int));}{\small \par}

~{\small{}~~~~...}{\small \par}

//~Return~the~local~copy

//~Free~its~memory~in~the~merge~function

{\small{}~~~~~}\textcolor{blue}{\small return~input\_cpy;}{\small \par}

{\small \}}{\small \par}

{\small int{*}~merge(int{*}~input1,~int~size1,~}{\small \par}

{\small{}~~int{*}~input2,~int~size2)~\{}{\small \par}

//~\textit{Intermediate~Step}

//~-{}-{}-{}-{}-{}-{}-{}-{}-{}-{}-{}-{}-{}-{}-{}-{}-

//~An~extra~scratch~array{\small{}~is~required~~~~~}{\small \par}

{\small{}~~~~~}\textcolor{blue}{\small int{*}~ext\_scr~=~new~int{[}size1+size2{]};}{\small \par}

{\small{}~~~~~...}{\small \par}

//~Free~the~memory~used~at~the~previous~level

{\small{}~~~~~}\textcolor{blue}{\small free(input1);~free(input2);}{\small \par}

//~Return~the~extra~scratch

//~Free~its~memory~at~the~next~level

{\small{}~~~~~}\textcolor{blue}{\small return~ext\_scr;}{\small \par}

//~-{}-{}-{}-{}-{}-{}-{}-{}-{}-{}-{}-{}-{}-{}-{}-{}-

{\small \}}{\small \par}

{\small{}~~~~~...}{\small \par}

{\small{}~~~~~else~if~(threads~>~1)~\{}{\small \par}

//~store~input\_copy~or~local~scratch~pointer~

{\small{}~~~~~~~~~int{*}~part1;~int{*}~part2;}{\small \par}

{\small \#pragma~omp~parallel~sections}{\small \par}

{\small{}~~~~~\{}{\small \par}

{\small \#pragma~omp~section}{\small \par}

{\small{}~~~~~~\{}{\small \par}

{\small{}~~~~~~~part1~=~mergesort\_parallel\_omp(input,~}{\small \par}

{\small{}~~~~~~~~~~~~~~~~~scratch,~size/2,~threads/2);~~~~~~~~~~~~~~~~~~~~}{\small \par}

{\small{}~~~~~~\}}{\small \par}

{\small \#pragma~omp~section}{\small \par}

{\small{}~~~~~~\{}{\small \par}

{\small{}~~~~~~~part2~=~mergesort\_parallel\_omp(input+size/2,}{\small \par}

{\small{}~~~~~~~~~~~~~~~~~scratch+size/2,~size-size/2,~}{\small \par}

{\small{}~~~~~~~~~~~~~~~~~threads-threads/2);~~~~~~~~~~~~~~~~~~~~}{\small \par}

{\small{}~~~~~~\}}{\small \par}

{\small{}~~~~~\}}{\small \par}

{\small{}~~~~~return~merge(part1,~size/2,~part2,~}{\small \par}

{\small{}~~~~~~~~~~~~~~size-size/2);}{\small \par}

{\small{}~~~~\}}{\small \par}

{\small{}~~return~(int{*})0;~}{\small \par}

{\small \}}{\small \par}

{\small int~main()~\{}{\small \par}

{\small{}~~~~...}{\small \par}

{\small{}~~~~int{*}~result~=~mergesort\_parallel\_omp(array0,~}{\small \par}

{\small{}~~~~~~scratch0,~ARRAY\_SZ,~NUM\_THREADS);}{\small \par}

{\small{}~~~~int{*}~temp~=~array0;}{\small \par}

{\small{}~~~~array0~=~result;}{\small \par}

{\small{}~~~~free(temp);}{\small \par}

{\small{}~~~~...}{\small \par}

{\small{}~~~~return~0;}{\small \par}

{\small \}}{\small \par}
\end{lyxcode}
\caption{\label{alg:Changes-to-code}Changes to code to \textit{localise} the
home caches }

\end{algorithm}

The most important feature of our approach is that we do not use architecture-specific
API. The Tilera platform is an ideal exploration platform because
of its fine-grained control over caching and memory, but we do not
expose any architecture-specific features to the programmer. We simply
advocate a different programming approach. Therefore, both  the original
and the \textit{localised} code can be run unmodified on any Linux
platform.

\section{Experiments and Discussion}

We have performed experiments for the different cases listed in Table
\ref{tab:Design-of-Experiments}. Although our approach has 3 building
blocks, in order to show the effect of each one independently, we
move smoothly from the conventional programming approach towards our
\textit{completely} \textit{localised} technique by changing one parameter
at a time (Case 1 to Case 8 in Table \ref{tab:Design-of-Experiments}).
However, we call any technique that copies the sub-arrays into dynamically
created arrays, a \textit{localised} technique (Case 5 to Case 8).
All cases are examined under memory striping mode, which balances
the memory traffic between all of the 4 memory controllers. Once again,
the default caching option of the system is to use \textit{hash for
home} for all types of user memory except the stack associated with
each task, which is homed locally on the task's current tile. We denote
this as \textit{all~but~stack}. Another caching policy which can
be used as a Tile Linux boot argument is not to use \textit{hash for
home} by default at all, which is homing the data on the tile that
is running the task, and is the desirable option for the \textit{localisation}
approach. We denote it as \textit{none} \cite{Tilera:Online}. We
also explore the effect of 2 mapping techniques: Decision of mapping
the threads to the cores can be leaved to the Tile Linux scheduler
or can be made statically as shown in the code (STATIC\_MAPPING).
Static mapping is performed by pinning each thread based on its thread
ID to the core with the same core ID. The efficiency of the method
described in Algorithm \ref{alg:Localisation} can be verified by
employing these 2 different mapping techniques.

\subsection{Evaluation of Speed-up }

\begin{figure}[t]
\begin{centering}
\includegraphics[width=0.9\columnwidth]{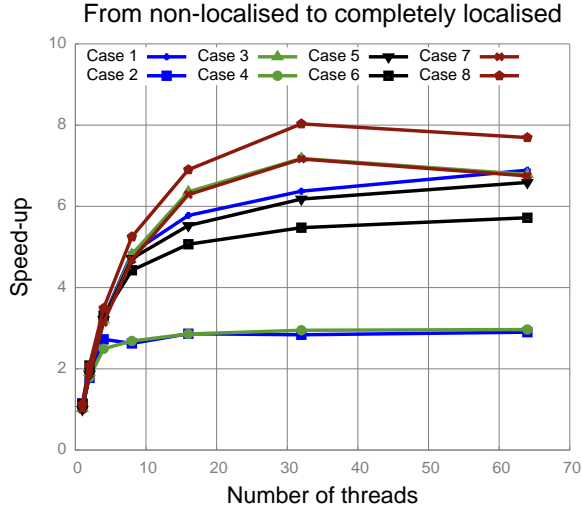}
\par\end{centering}

\caption{\label{fig:Speedup-cent-vs-decent}Speed-up of merge sort on 100M
integers on the TILEPro64 with memory striping enabled}
\end{figure}

Based on Figure \ref{fig:Speedup-cent-vs-decent}, it is evident that
the native GNU/Linux thread scheduling is not as efficient as static
mapping policy. This means that as long as the number of threads is
less than the number of cores, pinning the threads statically to the
processing cores and dividing the workload between them can result
in better performance. This way, high-cost thread migrations do not
occur multiple times during the execution time. The speed-up for all
the cases is computed against a base execution time, which is the
execution time with a single thread under the default hashing scheme
and the default Linux scheduling policy.

Figure \ref{fig:Speedup-cent-vs-decent} shows that in almost all
the cases, static mapping is the winning policy. Moreover, the \emph{localised}
approach significantly outperforms the \textit{non-localised} one
under the \textit{local homing}, whereas it does not lose the competition
under the \textit{hash for home} policy. Therefore, the three best
cases are Case 8, Case 7, and Case 3, all of them utilising the static
mapping strategy. We can get the best performance for this array computation
algorithm with complete \textit{localisation} under the \textit{local
homing} policy, which serves as a validation for our approach.

\begin{table}
\begin{tabular}{|c|c|c|c|c|}
\hline
{\bf Case} & {\bf Policy} & {\bf Mapper} & {\bf Hash} \tabularnewline \hline \hline
Case 1: & Non-localised & Tile Linux & All~but~stack  \tabularnewline \hline
Case 2: & Non-localised & Tile Linux & None  \tabularnewline \hline
Case 3: & Non-localised & Static Mapper & All~but~stack  \tabularnewline \hline
Case 4: & Non-localised & Static Mapper & None  \tabularnewline \hline
Case 5: & Localised & Tile Linux  & All~but~stack \tabularnewline \hline
Case 6: & Localised & Tile Linux & None  \tabularnewline \hline
Case 7: & Localised & Static Mapper & All~but~stack  \tabularnewline \hline
Case 8: & Localised & Static Mapper & None  \tabularnewline \hline
\hline  
\end{tabular}

\caption{\label{tab:Design-of-Experiments}Design of Experiments}
\end{table}

\subsection{Effect of Data Sizes\label{Effect-of-Data}}

As mentioned before, in order to isolate the effect of changes to
the conventional code, we have used an \textit{intermediate step},
which is marked on Algorithm \ref{alg:Changes-to-code}. This step
is in fact needed for the \textit{localised} style to work correctly,
but as a standalone technique, can optimise the parallel merge sort
algorithm by removing the need for copying back the merged array to
the inputs of the \textit{merge} function. In the conventional code,
the inputs of the \textit{merge} function are continuous parts of
the initial array, whereas in the new code, they are two copies that
are not allocated continuously in the memory. The only way to return
the merged result of them is to create an extra local scratch (\textit{ext\_scr})
inside the merge function and free it at the next level of the reduction
tree as the same as the \textit{input\_cpy} arrays are freed.

\begin{figure}[t]
\begin{centering}
\includegraphics[width=0.9\columnwidth]{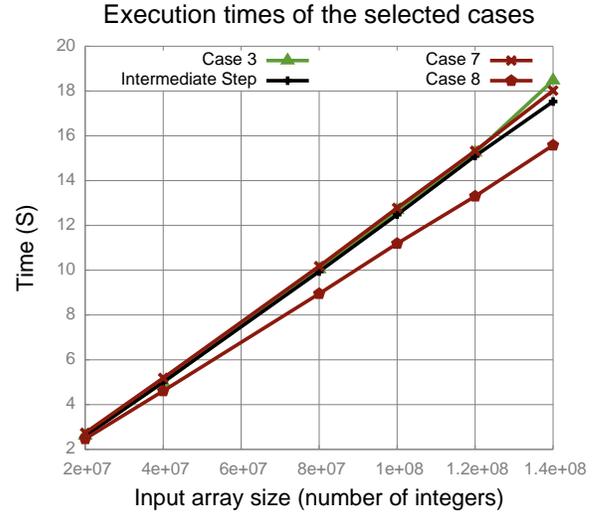}
\par\end{centering}

\caption{\label{fig:Best_times}Execution times of the best cases for different
input sizes}
\end{figure}

We use the Figure \ref{fig:Best_times} to show that, although this
technique results in slightly better performance, it does not have
any considerable influence compared to the \textit{localisation} method,
which is the main reason for getting a better execution time. All
of the cases in Figure \ref{fig:Best_times} are tested with 64 threads
on different input array sizes, under static mapping strategy with
memory striping enabled. Case 8 in this figure is the only one that
employs the \textit{local homing} policy. The \textit{intermediate
step} has a poor performance (close to that of Case 4) for the \textit{local
homing} policy, and its result is not included in the graph. Unlike
the \textit{input\_cpy}, the \textit{ext\_scr} cannot benefit from
the \textit{local homing}. The reason is that with maximum 64 threads,
most of the time is still spent on the \textit{mergesort\_serial}
function. The \textit{input\_cpy} array can benefit from recursive
access to its data which is cached on the local tile, while the merge
operation is performed only once and the \textit{ext\_scr} array is
freed afterward.

Figure \ref{fig:Best_times} shows that when the size of the input
array becomes larger, the \textit{localisation} style under the \textit{local
homing} policy can benefit more and more from data localisation, and
would show better performance than any programming style under the
\textit{hash for home} policy, either \textit{non-localised} or \textit{localised}.

\subsection{Effect of Memory Striping}

Memory striping, the automatic distribution of memory accesses over
all memory controllers, is the default behaviour of the TILEPro64.
Memory pages can be allocated either through a specific memory controller
or in striping mode, where each page is striped across all memory
controllers in 8KB chunks. With memory striping, Linux will boot up
believing it has a single memory controller that is four times larger
than any of the actual physical memory controllers. 

Static mapping might be criticised because of its poor behaviour in
utilisation of the memory controllers. In this work, we deliberately
mapped the threads to cores in an ordered way to show that thanks
to the caches, in either striping or non-striping mode, static mapping
outperforms the native GNU/Linux scheduler.

\begin{figure}[t]
\begin{centering}
\includegraphics[width=0.9\columnwidth]{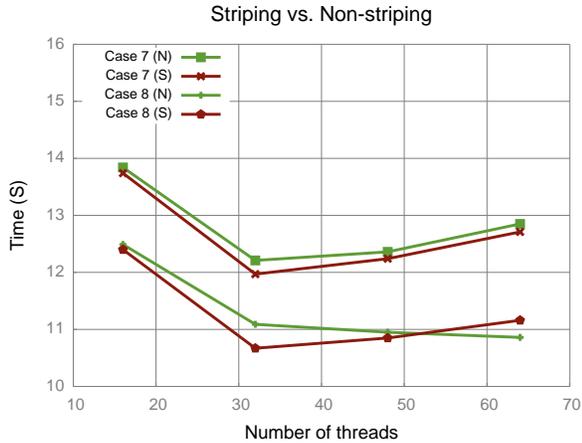}
\par\end{centering}

\caption{\label{fig:Effect-of-striping}Static mapping -- Influence of memory
striping on the execution time}
\end{figure}

In both cases, when moving from 16 to 32 threads, striping leads to
better performance. The reason is that with static mapping, all of
the threads are mapped to the upper rows of the chip (cores 0 to 31).
Therefore, in non-striping mode, they can only utilise 2 memory controllers.
Instead, in the striping mode, they can use all 4 controllers. From
32 to 64 threads, every new thread is mapped to the lower part of
the chip. Thus, with 64 threads, all 4 memory controllers are fully
utilised. In comparison with 32 threads, this makes the situation
better for non-striping mode, which can utilise more controllers,
and worse for striping mode, which experiences higher contention on
memory controllers. This behaviour is much more observable if one
turns off the caches. Nonetheless, in this application, when caching
is enabled, in contrast to the homing policy or mapping strategy,
memory striping does not show a significant effect. 

In summary, it is evident that in the traditional style, \textit{hash
for home} provides a much better performance. By contrast, in our
approach we do not specify either the \textit{hash for home} or \textit{local
homing} strategy of the Tilera Tile Processor to achieve better performance.
Rather, we present a programming technique which performs very well
in both cases. In addition, due to the localisation of the home caches
at a relatively coarse granularity, its performance is significantly
better when the \textit{local homing} strategy is used by the system's
hypervisor. The effect of memory striping is considerable when caching
is turned off across the system. However, when cashing is enabled,
it is mostly transparent to the user.

\section{Conclusion}

The use of a distributed shared caching hierarchy is a necessity in
emerging manycore systems. Improving the data locality is key to optimising
performance. It is important that the manycore programmer knows how
different caching mechanisms work. There is a lot of fine-grained
architecture-specific control that every new platform offer to its
customers, but in general, to benefit from these features, existing
codes have to be changed significantly in order to utilise the chip
efficiently. In this paper, we have introduced a programming approach
that is applicable to all manycore architectures with home caches. 

Our preliminary results show that our novel approach can outperform
the conventional programming style under the SMP Linux scheduler in
multithreaded environments. We also conclude that any \textit{hash
for home} policy, if used at the cache line granularity is too fine-grained
for parallel array computations with lots of sequential memory accesses.
Considering the benefits already obtained from this early stage work,
we believe that further research will lead to significant improvements
in the field of manycore programming. 

\bibliographystyle{acm}
\bibliography{heart2013}

\begin{thebibliography}{1}

\bibitem{bell2008tile64}
{\sc Bell, S., Edwards, B., Amann, J., Conlin, R., Joyce, K., Leung, V.,
  MacKay, J., Reif, M., Bao, L., Brown, J., et~al.}
\newblock Tile64-processor: A 64-core soc with mesh interconnect.
\newblock In {\em Solid-State Circuits Conference, 2008. ISSCC 2008. Digest of
  Technical Papers. IEEE International\/} (2008), IEEE, pp.~88--598.

\bibitem{kim2002adaptive}
{\sc Kim, C., Burger, D., and Keckler, S.~W.}
\newblock An adaptive, non-uniform cache structure for wire-delay dominated
  on-chip caches.
\newblock In {\em Acm Sigplan Notices\/} (2002), vol.~37, ACM, pp.~211--222.

\bibitem{morari2012efficient}
{\sc Morari, A., Tumeo, A., Villa, O., Secchi, S., and Valero, M.}
\newblock Efficient sorting on the tilera manycore architecture.
\newblock In {\em Computer Architecture and High Performance Computing
  (SBAC-PAD), 2012 IEEE 24th International Symposium on\/} (2012), IEEE,
  pp.~171--178.

\bibitem{muddukrishna2013task}
{\sc Muddukrishna, A., Podobas, A., Brorsson, M., and Vlassov, V.}
\newblock Task scheduling on manycore processors with home caches.
\newblock In {\em Euro-Par 2012: Parallel Processing Workshops\/} (2013),
  Springer, pp.~357--367.

\bibitem{radenskishared}
{\sc Radenski, A.}
\newblock Shared memory, message passing, and hybrid merge sorts for standalone
  and clustered smps.
\newblock In {\em Proc. PDPTA'11\/} (2011), CSREA press, pp.~367--373.

\bibitem{shim2012judicious}
{\sc Shim, K.~S., Lis, M., Khan, O., and Devadas, S.}
\newblock Judicious thread migration when accessing distributed shared caches.

\bibitem{Tilera:Online}
{\sc Tilera}.
\newblock Tile processor user architecture manual ug105, 2011.

\bibitem{tousimojarad2013parallel}
{\sc Tousimojarad, A.}
\newblock A parallel task composition approach to manycore programming.
\newblock {\em PLACES 2013\/} (2013), 29.

\end{thebibliography}

\end{document}